\documentclass[]{iopart}
\usepackage{graphicx}
\usepackage{iopams}

\begin{document}
\title[Entropy of entanglement in continuous frequency space of the biphoton state]
{Entropy of entanglement in continuous frequency space of the biphoton state from multiplexed cold atomic ensembles}
\author{H. H. Jen}
\address{Institute of Physics, Academia Sinica, Taipei 11529, Taiwan, R. O. C.}
\ead{sappyjen@gmail.com}
\renewcommand{\k}{\mathbf{k}}
\renewcommand{\r}{\mathbf{r}}
\newcommand{\f}{\mathbf{f}}

\begin{abstract}
We consider a scheme of multiplexed cold atomic ensembles that generate a frequency-entangled biphoton state with controllable entropy of entanglement.\ The biphoton state consists of a telecommunication photon (signal) immediately followed by an infrared one (idler) via four-wave mixing with two classical pump fields.\ Multiplexing the atomic ensembles with frequency and phase-shifted signal and idler emissions, we can manipulate and control the spectral property of the biphoton state.\ Mapping out the entropy of entanglement in the scheme provides the optimal configuration for entanglement resources.\ This paves the way for efficient long-distance quantum communication and for potentially useful multimode structures in quantum information processing.
\end{abstract}
\pacs{42.50.Dv, 03.67.Bg, 03.67.Hk}
\submitto{\jpb}
\maketitle
\section{Introduction}
Quantum communication relies on the coherent distribution of entanglement over long distances.\ This can be done by implementing a quantum repeater protocol \cite{Briegel1998, Dur1999}.\ It distributes the entanglement to the end parties (A and B) of distance L by inserting M initially entangled pairs (A and C$_1'$, C$_2$ and C$_2'$,..., C$_{\rm M}$ and B) respectively with a distance L/M.\ Conditioning on locally joint measurements (e.g. Bell measurements) of the adjacent parties (C$_i'$ and C$_{i+1}$), the protocol succeeds and projects out the required entangled state of the end parties with some finite fidelity $F$ $\leq$ $1$.\ Since then the long-distance quantum communication has been proposed in the setup of atomic ensembles \cite{Duan2001} where entanglement swapping or quantum teleportation \cite{Pirandola2015} becomes feasible.\ In about the last decade, correlated atom-light entanglement in the Raman-type \cite{Matsukevich2004, Chou2004, Chaneliere2005, Chen2006, Laurat2006} and diamond-type schemes \cite{Chaneliere2006, Radnaev2010, Jen2010, Jen2012-2} paves the way toward the realization of low-loss long-distance quantum communication.
%%%%%%%%%%%%%%%%%%%%%%%%%%%%%%%%%%%%%%%%%%%%%

The entanglement is a basic element in long-distance quantum communication, which has been focused on discrete degrees of freedom either in light polarizations \cite{Clauser1969, Aspect1981, Kwiat1995} or central frequencies \cite{Lan2007, Ramelow2009}.\ Recently the entanglement of continuous variables provides a richer capacity in quantum key distributions \cite{Gisin2002} and quantum information applications \cite{Braunstein2005}.\ A plethora of continuous degrees of freedom involve light spectrum \cite{Branning1999, Law2000, Parker2000}, transverse momentum \cite{Law2004, Moreau2014}, space \cite{Grad2012}, and orbital angular momenta of light \cite{Arnaut2000, Mair2001, Molina2007, Dada2011, Agnew2011,Fickler2012}.\ This higher dimensional quantum capacity also manifests in the aspects of quantum memories using atomic ensembles \cite{Nicolas2014, Ding2015}, and in proposed atomic \cite{Afzelius2009} or optical frequency comb techniques \cite{Zheng2015}.\ In addition, multiplexing multimode quantum memories in space \cite{Collins2007, Lan2009} or time \cite{Simon2007} can enhance the distribution rate of quantum repeater protocol while the spectral shaping in spontaneous parametric down conversion \cite{Bernhard2013, Lukens2014} or diamond-type atomic ensemble \cite{Jen2015-2} helps facilitate the frequency encoding/decoding, which promises a potentially efficient multimode quantum communication. 
%%%%%%%%%%%%%%%%%%%%%%%%%%%%%%%%%%%%%%%%%%%%%

This motivates us to investigate the entropy of entanglement in continuous frequency space of the biphoton state from the multiplexed cold atomic ensembles where its telecommunication (telecom) bandwidth has the advantage of low-loss fiber transmission while its spectrum can be manipulated in the multiplexed scheme.\ Based on the studies of spectrally-entangled cascade emissions \cite{Jen2012-2} and their spectral shaping via frequency shifters in the multiplexed atomic ensembles \cite{Jen2015-2}, we further propose to manipulate the spectral property with additional phase shifters in the multiplexed scheme.\ Other than previous focus on DLCZ (Duan-Lukin-Cirac-Zoller) protocol \cite{Duan2001} utilizing spectrally-entangled photons \cite{Jen2012-2}, here we extend the multiplexed scheme \cite{Jen2015-2} by including the degree of freedom of phases in the biphoton state, which offers more flexible control over its entanglement property.\ In this article, we first introduce the biphoton state in continuous frequency space generated from driving a cold atomic ensemble via four-wave mixing in section 2.\ In section 3, we consider a scheme of multiplexed cold atomic ensembles via frequency and phase shifters, that can provide flexibility and controllability over the spectral property of the biphoton state.\ The entropy of entanglement in our scheme is characterized in details up to three atomic ensembles in section 4.\ We then conclude in section 5 and discuss the potential applications of our scheme in multimode long-distance quantum communication.
%%%%%%%%%%%%%%%%%%%%%%%%%%%%%%%%%%%%%%%%%%%%%%%%%%%%%%%%%%%%%%
\section{Biphoton state in continuous frequency space}
We consider the biphoton state generated by a cold Rb atomic ensemble with a diamond-type atomic level in figure \ref{fig1}.\ Two classical driving fields operate on the the infrared and telecom transitions, and then within four-wave mixing condition, correlated signal and idler photons in telecom and infrared bandwidths respectively are spontaneously emitted.\ The Hamiltonian and the coupled equations in Schr\"odinger picture has been derived and solved \cite{Jen2012-2} where we formulate the light-atom interactions with the dipole approximation \cite {QO:Scully}.\ The adiabatic approximation is assumed to be valid if the Rabi frequencies of the pump fields $\Omega_a$, $\Omega_b$ are weak enough with large detunings $\Delta_1$ $\equiv$ $\omega_a-\omega_1$, $\Delta_2$ $\equiv$ $\omega_a$ $+$ $\omega_b$ $-$ $\omega_2$ where $\omega_{1,2}$ (and $\omega_3$) are atomic level energies.\ The main result is the biphoton probability amplitude \cite{Jen2012-2}
\begin{eqnarray}\fl
D_{si}(\Delta\omega_s,\Delta\omega_i)=\frac{\tilde{\Omega}_a\tilde{\Omega}_b g_s^*g_i^*(\epsilon_{s}^*\cdot\hat{d}_s)(\epsilon_{i}^*\cdot\hat{d}_i)\sum_\mu e^{i\Delta\k\cdot\r_\mu}}{4\Delta_1\Delta_2\sqrt{2\pi}\tau}f(\omega_s,\omega_i),\label{Dsi}
\end{eqnarray}
where the spectral function of the biphoton state is
\begin{eqnarray}
f(\omega_s,\omega_i)=\frac{e^{-(\Delta\omega_s+\Delta\omega_i)^2\tau^2/8}}{\frac{\Gamma_3^{\rm N}}{2}-i\Delta\omega_i}.
\end{eqnarray}
The pump fields are normalized Gaussian pulses $\Omega_{a,b}(t)$ $=$ $\frac{1}{\sqrt{\pi}\tau}\tilde{\Omega}_{a,b}e^{-t^2/\tau}$ where $\tilde{\Omega}_{a,b}$ is the pulse area.\ The coupling constants of signal and idler photons are $g_{s,i}$ with polarizations $\epsilon_{s,i}$ and unit direction of dipole moments $\hat{d}_{s,i}$.\ The four-wave mixing (FWM) condition for two pump fields (of wavevectors $\k_{a,b}$) and two photons ($\k_{s,i}$) is $\Delta\k$ $=$ $\k_a$ $+$ $\k_b$ $-$ $\k_s$ $-$ $\k_i$, and $\Delta\omega_s$ $\equiv$ $\omega_s$ $-$ $\omega_{23}$ $-$ $\Delta_2$ $-$ $\delta\omega_i$, $\Delta\omega_i$ $\equiv$ $\omega_i$ $-$ $\omega_3$ $+$ $\delta\omega_i$ with central frequencies of light $\omega_{s,i}$.\ The telecom transition frequency is $\omega_{23}$ $\equiv$ $\omega_2$ $-$ $\omega_3$, and $\omega_3$ is the infrared one.\ The spectral function shows a Gaussian envelope that maximizes at energy conservation of two photons $\Delta\omega_s$ $+$ $\Delta\omega_i$ $=$ $0$, which is modulated by a Lorentzian of an idler photon.\ The FWM condition guarantees the generation of directionally correlated signal and idler photons, and selectively drives the atomic system into a symmetrically collective excitation \cite{Eberly2006, Scully2006, Mazets2007}.\ In such atomic system with an atomic density $\sim$ $10^{11}$ cm$^{-3}$, the idler photon is observed to be superradiant \cite{Chaneliere2006, Jen2012, Srivathsan2014} due to induced dipole-dipole interactions \cite{Lehmberg1970} firstly proposed as Dicke's radiation \cite{Dicke1954}.\ The superradiant decay constant is denoted as $\Gamma_3^{\rm N}$ \cite{Rehler1971} and $\delta\omega_i$ is the associated cooperative Lamb shift \cite{Friedberg1973, Scully2009, Jen2015}.

The four atomic levels can be chosen as ($|0\rangle$, $|1\rangle$, $|2\rangle$, $|3\rangle$) $=$ ($|5\textrm{S}_{1/2},\textrm{F}=3\rangle$, $|5\textrm{P}_{3/2},\textrm{F}=4\rangle$, $|4\textrm{D}_{5/2},\textrm{F}=5\rangle$, $|5\textrm{P}_{3/2},\textrm{F}=4\rangle$) where $|2\rangle$ could be also 6S$_{1/2}$, 7S$_{1/2}$, or 4D$_{3/2}$ that the telecom bandwidth resides in 1.3-1.5 $\mu$m \cite{Chaneliere2006}.\ The spectral property of this biphoton state can be analyzed by Schmidt decomposition \cite{Law2000}, where the state vector can be expressed in terms of Schmidt eigenvalues $\lambda_n$ and effective photon operators $\hat{b}_n$, $\hat{c}_n$ that the biphoton state becomes
\begin{eqnarray}
|\Psi\rangle &=& \int f(\omega_s,\omega_i)\hat{a}_{\lambda_s}^\dag(\omega_s)\hat{a}_{\lambda_i}^\dag(\omega_i)|0\rangle d\omega_s d\omega_i,\label{psi}\\
&=&\sum_{n}\sqrt{\lambda_n}\hat{b}_n^\dag\hat{c}_n^\dag|0\rangle,
\end{eqnarray}
with signal and idler mode functions $\psi_n$, $\phi_n$ that define the effective photon operators,
\begin{eqnarray}
\hat{b}_n^\dag&\equiv&\int\psi_n(\omega_s)\hat{a}_{\lambda_s}^\dag(\omega_s)d\omega_s,\\
\hat{c}_n^\dag&\equiv&\int\phi_n(\omega_i)\hat{a}_{\lambda_i}^\dag(\omega_i)d\omega_i.
\end{eqnarray}
Furthermore the entropy of entanglement can be written as
\begin{eqnarray}
S=-\sum_{n=1}^\infty\lambda_n\log_{2}\lambda_n.
\end{eqnarray}

This biphoton state in continuous frequency space can be implemented in the DLCZ (Duan-Lukin-Cirac-Zoller) protocol \cite{Duan2001}, that is advantageous for its telecom bandwidth for low-loss fiber transmission.\ Combined with its infrared bandwidth suitable for the quantum storage locally, the entanglement swapping and quantum teleportation have been investigated using such cascade emissions \cite{Jen2012-2}.\ In the next section, we consider a scheme of multiplexed atomic ensembles that would allow us to manipulate and control the entropy of entanglement.\ In addition we may also control the mode functions of the biphoton state.
%%%%%%%%%%%%%%%%%%%%%%%%%%%%%%%%%%%%%%%%%%%%%%%%%%%%%%%%%%%%%%
\section{Multiplexed cold atomic ensembles}
%%%%%%%%%%%%%%%%%%%%%%%%%%%%%%%%%%%%%%%%%%%%%%%%%%%%%%%%%%%%%%
\begin{figure}[t]
\begin{center}
\includegraphics[height=7.5cm, width=10cm]{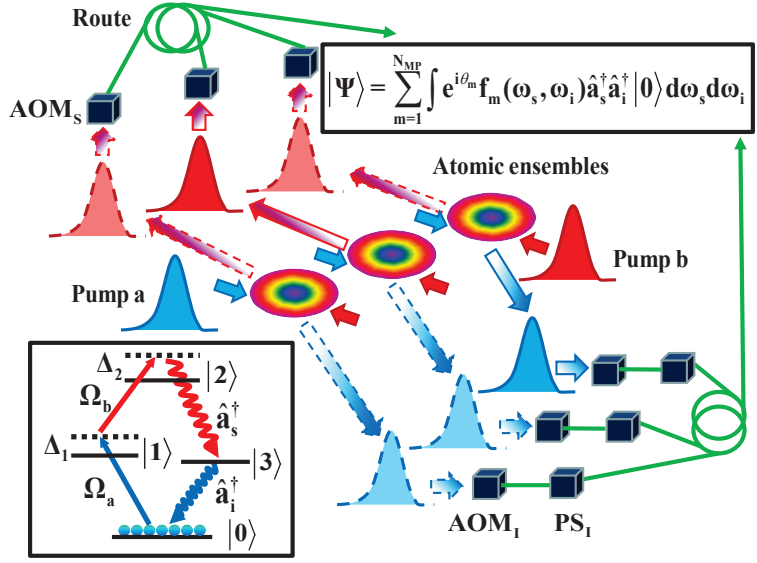}
\caption{Schematic multiplexed cold atomic ensembles, and diamond-type atomic level structure.\ The biphoton state of signal and idler photons $\hat{a}^\dagger_{s,i}$ is generated by two pump fields $\Omega_{a,b}$ with single and two-photon detunings $\Delta_{1,2}$.\ The circle represents a routing fiber coupler after acoustic-optic modulators (AOM) and phase shifters (PS) of the cascade-emitted photon pair (arrows).\ The effective biphoton state $|\Psi\rangle$ is derived where N$_{\rm MP}$ is the total number of atomic ensembles and $\theta_{m}$ is the phase shift by PS respectively for $m$th atomic ensemble.\ Here for demonstration we plot three atomic ensembles, and the dash waveforms of signal and idler photons represent the probabilistic nature of the multiplexed biphoton state.\ Pump a (b) and idler (signal) photon are denoted as blue (red) in the online version of color scheme.}\label{fig1}
\end{center}\end{figure}
%%%%%%%%%%%%%%%%%%%%%%%%%%%%%%%%%%%%%%%%%%%%%%%%%%%%%%%%%%%%%%
Here we consider a scheme of the multiplexed atomic ensembles as shown in figure \ref{fig1}.\ We excite the atomic ensembles with common pump fields, and the spontaneously emitted cascade emissions are multiplexed by the frequency and phase shifters that may control the central frequencies and relative phases of photons.\ The frequency shifts can be done by acoustic-optic modulators while the phase shifters can be implemented as in cross-phase modulation experiments by electromagnetically-induced-transparency based Kerr medium \cite{Kang2003}, atomic ensembles \cite{Lo2010, Shiau2011}, or hollow-core photonic bandgap fiber \cite{Venkataraman2013}, in the low light level.\ Similar to DLCZ protocol where the excitation probability is made small that multiphoton events are rare \cite{Duan2001}, we may also express the multiplexed cascade emissions as product states of the biphoton state in equation (\ref{psi}).\ Along with a prefactor in equation (\ref{Dsi}) where we denote as probability $p$ $\equiv$ $D_{s,i}(\Delta\omega_s,\Delta\omega_i)/f(\omega_s,\omega_i)$ that depends only on excitation parameters, we have
\begin{eqnarray}\fl
|\Psi\rangle_{\rm MP}=(|0\rangle_1+p|\Psi\rangle_1)\otimes(|0\rangle_2+p|\Psi\rangle_2)\otimes \dots\otimes(|0\rangle_{\rm N_{\rm MP}}+p|\Psi\rangle_{\rm N_{\rm MP}}),
\end{eqnarray}
where $|0\rangle$ means the ground state, and the subscripts denote the numbered atomic ensembles.\ We note that these atomic ensembles share the same excitation probability due to the common pump fields we apply, which furthermore remove the incoherent (random) relative phases that may deteriorate the state preparation in our scheme.\ After expanding the above product states and keeping the first two most significant terms, we derive
\begin{eqnarray}
|\Psi\rangle_{\rm MP}=|0\rangle^{\otimes \rm N_{\rm MP}} + pf_{\rm MP}(\omega_s,\omega_i) + \mathcal{O}(p^2),
\end{eqnarray}
where the events of more than two photons are in the order of $p^2$ which is extremely small in our assumption of weak excitations.
%%%%%%%%%%%%%%%%%%%%%%%%%%%%%%%%%%%%%%%%%%%%%%%%%%%%%%%%%%%%%%

Therefore the spectral function of the effective multiplexed biphoton state can be written as 
\begin{eqnarray}
f_{\rm MP}(\omega_s,\omega_i)=\sum_{m=1}^{\rm N_{\rm MP}}e^{i\theta_m}\frac{e^{-(\Delta\omega_s+\Delta\omega_i+\delta q_m)^2\tau^2/8}}{\frac{\Gamma_3^N}{2}-i(\Delta\omega_i+\delta p_m)},
\end{eqnarray}
where N$_{\rm MP}$ is the number of the multiplexed atomic ensembles, $\delta p_m$ and $\delta q_m$ are frequency shifts respectively for idler and jointly signal and idler photons.\ Phase shift is denoted as $\theta_m$ which can be addressed independently for each atomic ensemble by phase shifters.
%%%%%%%%%%%%%%%%%%%%%%%%%%%%%%%%%%%%%%%%%%%%%%%%%%%%%%%%%%%%%%

The spectral shaping for the above multiplexed scheme has been investigated for symmetrical ($\delta q_m$ $=$ $0$) and nonsymmetric ($\delta q_m$ $=$ $\delta p_m$) spectral functions \cite{Jen2015-2}.\ In the symmetrical spectral function, the spectral weighting lies along the energy conserving axis $\Delta\omega_s$ $=$ $-\Delta\omega_i$, which generates a larger entropy of entanglement S.\ Here we add another degree of freedom in the phase of the photons, that provides richer information in S and more flexibility of controlling it.
%%%%%%%%%%%%%%%%%%%%%%%%%%%%%%%%%%%%%%%%%%%%%%%%%%%%%%%%%%%%%%
\section{Entropy of entanglement}
%%%%%%%%%%%%%%%%%%%%%%%%%%%%%%%%%%%%%%%%%%%%%%%%%%%%%%%%%%%%%%
\begin{figure}[t]
\begin{center}
\includegraphics[height=5.0cm, width=10cm]{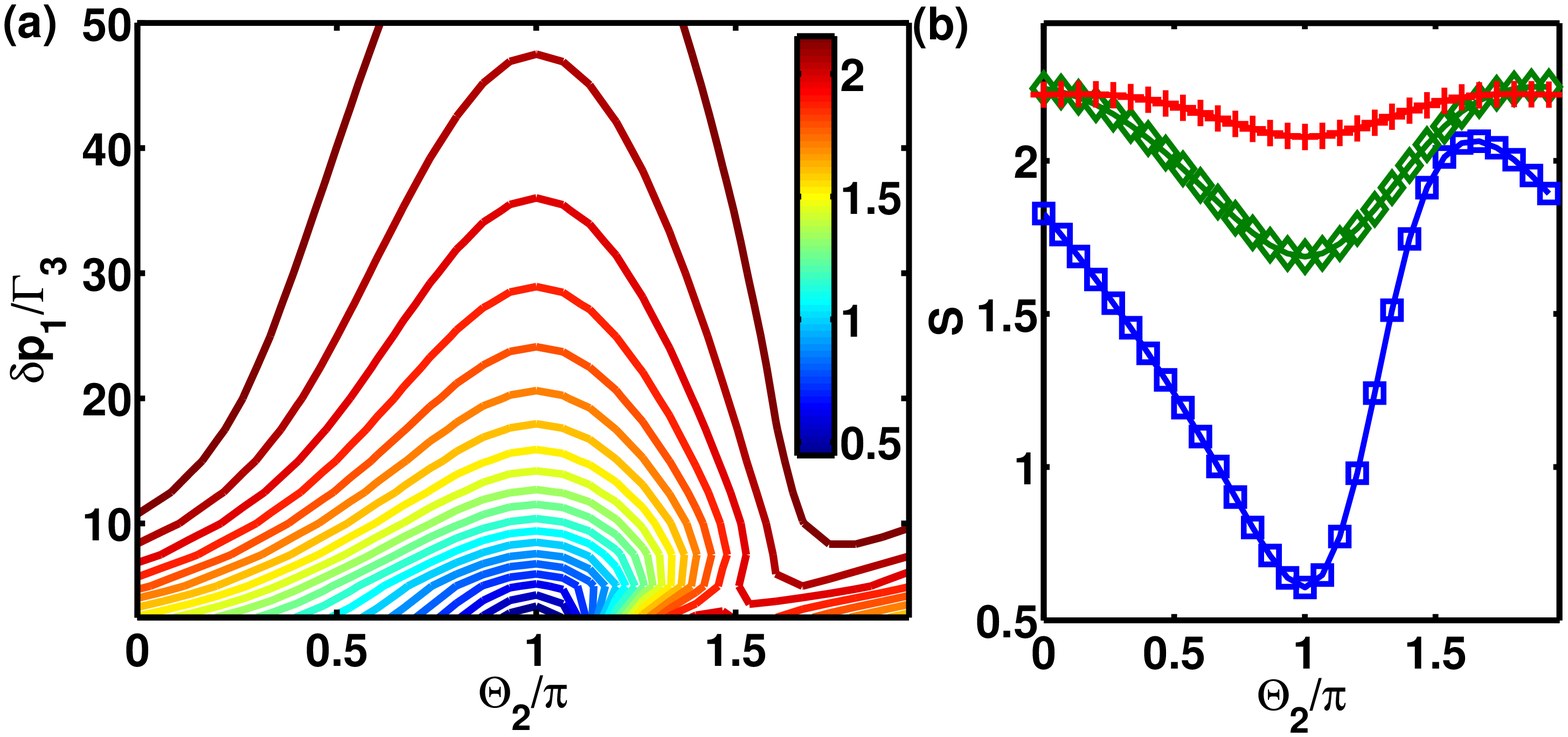}
\caption{Entropy of entanglement S for two atomic ensembles.\ (a) Contour plot of S in frequency $\delta p_1$ and phase shifts $\theta_2$ with the symmetrical spectral function set as $\delta p_2$ $=$ $-\delta p_1$ and $\delta q_{1,2}$ $=$ $0$. (b) Sectional plots of (a) at $\delta p_1/\Gamma_3$ $=$ $5$ ($\square$), $20$ ($\diamond$), and $50$ ($+$), which approach a plateau as $\delta p_1$ increases.\ The spectral ranges for both signal and idler photons are
post-selected to $\pm$ $300$ $\Gamma_3$ in all figures where we also set $\Gamma_3^{\rm N}/\Gamma_3$ $=$ $5$ and $\tau$ $=$ $0.25\Gamma_3^{-1}$ without loss of generality.}\label{fig2}
\end{center}
\end{figure}
%%%%%%%%%%%%%%%%%%%%%%%%%%%%%%%%%%%%%%%%%%%%%%%%%%%%%%%%%%%%%%
The entropy of entanglement S is crucial in evaluating the capacity of Hilbert space that the quantum system can access.\ In the setting of discrete quantum system, for a maximally entangled qudit state of dimensions D (or W state), we have
\begin{eqnarray}
|\Psi\rangle_{\rm D}=\frac{1}{\sqrt{\rm D}}(|100\dots 00\rangle+|010\dots 00\rangle+\dots+|000\dots 01\rangle),
\end{eqnarray}
where $|0\rangle$ and $|1\rangle$ represent a qubit space (e.g. polarizations for a single photon).\ The Hilbert space would involve $2^{D}$ states while the entropy of entanglement S becomes log$_2{\rm D}$.\ Therefore quantifying S allows us to analyze the capacity for quantum information processing and application.\ Other than discrete quantum system, here we focus on the capacity for long-distance quantum communication in our multiplexed scheme in continuous frequency space.\ There is no general analytical expression to S in continuous space, which however can be quantified using Schmidt decomposition.\ In addition, we later will show in equation (\ref{eq14}) that S in our multiplexed scheme can be approximately expressed as a summation of the entropy for qudit state of dimensions N$_{\rm MP}$ and the one in continuous frequency space with N$_{\rm MP}$ $=$ $1$.\ Below we consider two and three atomic ensembles, and we map out the entropy of entanglement in the setting of symmetrical spectral functions and also the associated mode probability densities for the biphoton state.
%%%%%%%%%%%%%%%%%%%%%%%%%%%%%%%%%%%%%%%%%%%%%%%%%%%%%%%%%%%%%%
\subsection{Multiplexed two atomic ensembles}
For the multiplexed scheme with two atomic ensembles in the symmetrical setting of frequency shifts where we set $\delta p_2$ $=$ $-\delta p_1$ with $\delta q_{1,2}$ $=$ $0$, the spectral function becomes
\begin{eqnarray}
f_{\rm MP}(\omega_s,\omega_i)=\frac{e^{-(\Delta\omega_s+\Delta\omega_i)^2\tau^2/8}}{\frac{\Gamma_3^N}{2}-i(\Delta\omega_i+\delta p_1)}
+\frac{e^{i\theta_2}e^{-(\Delta\omega_s+\Delta\omega_i)^2\tau^2/8}}{\frac{\Gamma_3^N}{2}-i(\Delta\omega_i-\delta p_1)},
\end{eqnarray}
where the phase $\theta_1$ is set to zero without loss of generality for the overall phase is irrelevant for the biphoton state.\ The entropy of entanglement S in such scheme is shown in figure \ref{fig2}.\ For larger frequency shifts, we have a plateau of entropy distribution where its maximum appears at $\theta_2$ $=$ $0$ while the minimum is at $\theta_2$ $=$ $\pi$.\ For smaller frequency shifts where multiplexed spectral functions start to overlap, the S has maximum near $\theta_2$ $=$ $2\pi$ while the minimum is still at phase $\pi$ which indicates an anti-symmetrical distribution to the axis of $\Delta\omega_s$ $=$ $\Delta\omega_i$.\ We note that as $\delta p_1$ approaches zero which indicates no frequency shifts to the atomic ensembles, the S at phase $\pi$ has no physical meaning since the spectral function becomes null.\ At $\delta p_1$ $=$ $0$, the spectral function has no difference from the one with just a single atomic ensemble for $\theta_2$ $\neq$ $\pi$.

%%%%%%%%%%%%%%%%%%%%%%%%%%%%%%%%%%%%%%%%%%%%%%%%%%%%%%%%%%%%%%
\begin{figure}[t]
\begin{center}
\includegraphics[height=5.0cm, width=10cm]{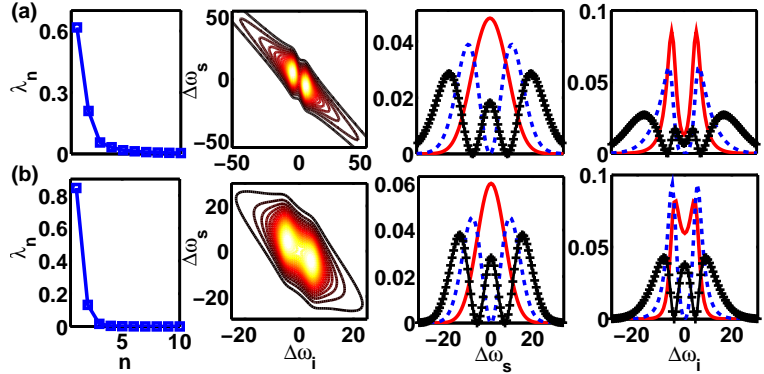}
\caption{Schmidt decomposition and mode probability densities for small frequency shifts.\ Schmidt eigenvalues and absolute spectral distribution are shown at $\delta p_1/\Gamma_3$ $=$ $5$ and $\theta_2$ $=$ $0$, $\pi$ respectively in (a) and (b).\ Associated first three, $n$ $=$ $1$, $2$, $3$ (solid, dash, and $+$), mode probability densities of signal $|\psi_n|^2$ and idler photons $|\phi_n|^2$ are put in horizontal orders respectively.}\label{fig3}
\end{center}\end{figure}
%%%%%%%%%%%%%%%%%%%%%%%%%%%%%%%%%%%%%%%%%%%%%%%%%%%%%%%%%%%%%%
Now we investigate in details of the eigenvalues and mode probability densities at some specific points in figure \ref{fig2}(a).\ In figure \ref{fig3} we study the spectral functions with small frequency shifts.\ The relative large and minimal entropy S at $\theta_2$ $=$ $0$ and $\pi$ respectively in (a) and (b) can be understood in the descending Schmidt eigenvalues where the largest eigenvalue closer to one in figure \ref{fig3}(b) means a less entangled biphoton state.\ In the extreme case when $\lambda_1$ $=$ $1$, the biphoton state becomes an unentangled source such that the signal and idler photons are separate in their respective mode functions.\ Zero phase in (a) with a larger S can be also seen in the spectral distribution that aligns on the axis $\Delta\omega_s$ $+$ $\Delta\omega_i$ $=$ $0$, in contrast to (b) that the phase $\pi$ tends to distribute the spectral weighting on $\Delta\omega_{s,i}$ $=$ $0$.\ We demonstrate three mode probability densities for this biphoton state where signal and idler photons show Gaussian and Lorentzian tails respectively as expected.\ The idler mode with the largest eigenvalue has double peaks, which would show interference patterns in time domains.\ This feature discriminates the characteristics of the signal mode of the largest eigenvalue from the idler one.
%%%%%%%%%%%%%%%%%%%%%%%%%%%%%%%%%%%%%%%%%%%%%%%%%%%%%%%%%%%%%%
\begin{figure}[t]
\begin{center}
\includegraphics[height=5.0cm, width=10cm]{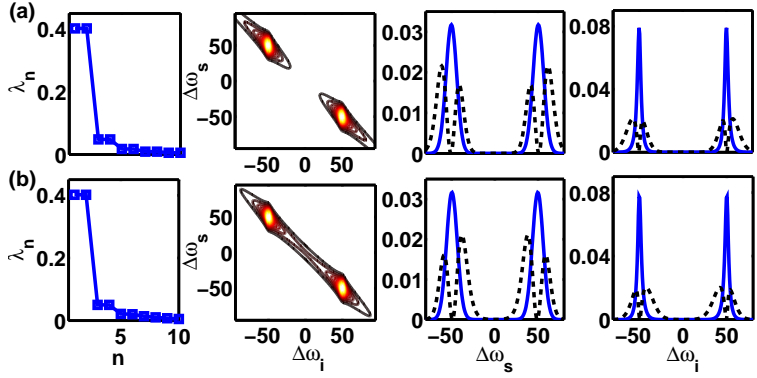}
\caption{Schmidt decomposition and mode probability densities for large frequency shifts.\ Schmidt eigenvalues and absolute spectral distribution are shown at $\delta p_1/\Gamma_3$ $=$ $50$ and $\theta_2$ $=$ $0$, $\pi$ respectively in (a) and (b).\ Associated first two degenerate, $n$ $=$ $1$ and $2$ (solid), and next two degenerate, $n$ $=$ $3$ and $4$ (dash) mode probability densities of signal $|\psi_n|^2$ and idler photons $|\phi_n|^2$ are plotted accordingly.}\label{fig4}
\end{center}\end{figure}
%%%%%%%%%%%%%%%%%%%%%%%%%%%%%%%%%%%%%%%%%%%%%%%%%%%%%%%%%%%%%%

In figure \ref{fig4}, similar to figure \ref{fig3}, we show the corresponding results for large frequency shifts.\ A pair of degenerate eigenvalues appear along with degenerate mode probability densities.\ For large $\delta p_1$, S deviates not much on $\theta_2$ as can be seen in figure \ref{fig2}(b) and also in the first ten eigenvalues in figure \ref{fig4} which can not be distinguished.\ The spectral distributions in figure \ref{fig4}(a) and (b) differ most in between these two spectral functions on $\Delta\omega_s$ $=$ $-\Delta\omega_i$, which reflects on the slightly different third and fourth mode probability densities though the largest two modes show no difference.\ The well-separated mode functions (in contrast to overlapped ones in figure \ref{fig3}) in frequency space provide the possibility to address and manipulate the frequency coding/encoding \cite{Bernhard2013}.\ Therefore the biphoton state in the multiplexed scheme can potentially implement the Hadamard codes \cite{Lukens2014}.

In the next subsection we further study three multiplexed atomic ensembles where we show how the proposed scheme offers complexity in spectral properties with frequency and phase shifts, and also the potentiality in multimode quantum information processing.
\subsection{Multiplexed three atomic ensembles}
For the multiplexed scheme with three atomic ensembles in the symmetrical setting of frequency shifts where we set $\delta p_2$ $=$ $-\delta p_1$, $\delta p_3$ $=$ $0$ with again $\delta q_{1,2,3}$ $=$ $0$, the spectral function becomes
\begin{eqnarray}\fl
f_{\rm MP}(\omega_s,\omega_i)=\frac{e^{i\theta_1}e^{-(\Delta\omega_s+\Delta\omega_i)^2\tau^2/8}}{\frac{\Gamma_3^N}{2}-i(\Delta\omega_i+\delta p_1)}
+\frac{e^{-(\Delta\omega_s+\Delta\omega_i)^2\tau^2/8}}{\frac{\Gamma_3^N}{2}-i\Delta\omega_i}
+\frac{e^{i\theta_2}e^{-(\Delta\omega_s+\Delta\omega_i)^2\tau^2/8}}{\frac{\Gamma_3^N}{2}-i(\Delta\omega_i-\delta p_1)},\label{three}
\end{eqnarray}
where again we set zero $\theta_3$ for irrelevant overall phase of the biphoton state.

%%%%%%%%%%%%%%%%%%%%%%%%%%%%%%%%%%%%%%%%%%%%%%%%%%%%%%%%%%%%%%
\begin{figure}[b]
\begin{center}
\includegraphics[height=5.0cm, width=10cm]{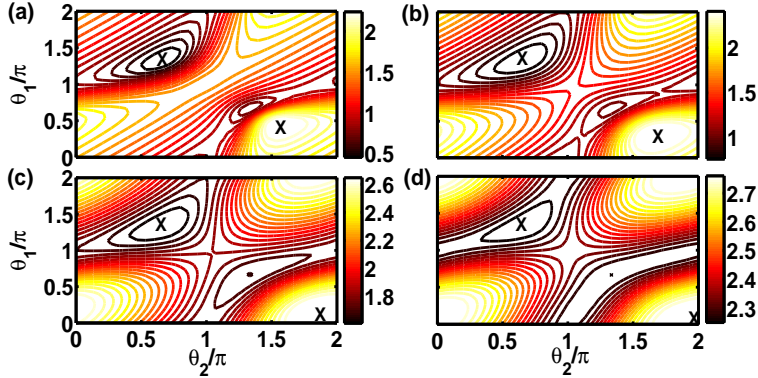}
\caption{Entropy of entanglement S for three atomic ensembles.\ Contour plots of S of the symmetrical spectral function in dimensions of $\theta_{1,2}$ with frequency shifts of $\delta p_1/\Gamma_3$ $=$ (a) $3$, (b) $6$, (c) $15$, and (d) $30$.\ The maximum of S approaches to $\theta_{1,2}$ $=$ $0$, $2\pi$ as $\delta p_1$ increases while its minimum fixes at $\theta_1$ $=$ $4\pi/3$ and $\theta_2$ $=$ $2\pi/3$.\ The crosses represent the extreme points in S.}\label{fig5}
\end{center}
\end{figure}
%%%%%%%%%%%%%%%%%%%%%%%%%%%%%%%%%%%%%%%%%%%%%%%%%%%%%%%%%%%%%%
In figure \ref{fig5} we map out the complete entropy of entanglement S for the biphoton state in equation (\ref{three}) from small to large frequency shifts.\ The maximum S approaches the four corners of the contour plots in phases, which are $\theta_{1,2}$ $=$ $0$, $2\pi$, as $\delta p_1$ increases.\ It also suggests that the S has relatively small deviations for larger frequency shifts similar to the case of two multiplexed atomic ensembles.\ The map of S indicates the asymmetry in two dimensions of $\theta_{1,2}$ with finite $\delta p_1$, and its minimum is observed to fix at $\theta_{1,2}$ $=$ $4\pi/3$ and $2\pi/3$ respectively.\ Furthermore even though the S can be manipulated to increase by increasing $\delta p_1$, from (a) to (c) we can see that the maximum S shows up at some optimal $\theta_{1,2}$ that offer more degrees of freedom to generate maximal S for some specific frequency shift.

\begin{figure}[b]
\begin{center}
\includegraphics[height=5.0cm, width=10cm]{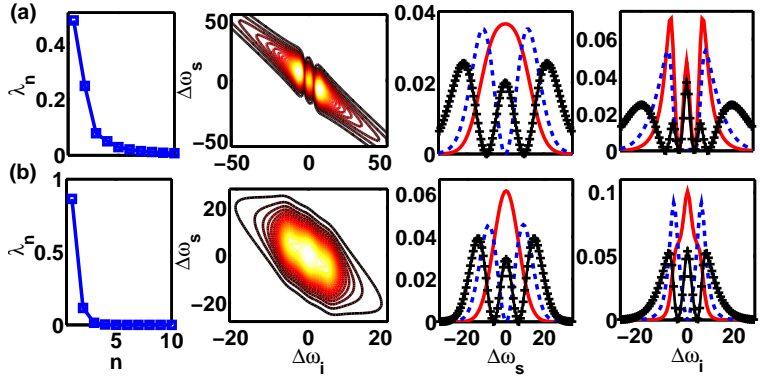}
\caption{Schmidt decomposition and mode probability densities for three multiplexed atomic ensembles.\ Schmidt eigenvalues and absolute spectral functions are shown at $\delta p_1/\Gamma_3$ $=$ $6$ with ($\theta_1$, $\theta_2$) $=$ ($\pi/3$, $5\pi/3$) and ($4\pi/3$, $2\pi/3$) in (a) and (b) respectively.\ Associated first three, $n$ $=$ $1$, $2$, and $3$ (solid, dash, and $+$) mode probability densities of signal $|\psi_n|^2$ and idler photons $|\phi_n|^2$ are plotted accordingly.}\label{fig6}
\end{center}\end{figure}

In figure \ref{fig6} we investigate closely on the extreme points of figure \ref{fig5}(b).\ Again the minimal S reflects on its descending eigenvalues where the largest eigenvalue is closer to one.\ Similar to two multiplexed atomic ensembles, the first idler mode in (a) has triple peaks due to three atomic ensembles being multiplexed.\ The large S in (a) also reflects on its spectral function that aligns mostly on the axis that conserves photon energies in contrast to the centrally distributed one in (b).\ We also find an interesting feature in the first idler mode of (b) which has two small humps around the central peak, and they grow up as S increases in the map of figure \ref{fig5}.\ For the first signal modes, we observe that the effect of narrowing in their linewidths is more significant by manipulating S compared to figure \ref{fig3} in the setting of two multiplexed atomic ensembles.\ In the scheme of multiplexed three atomic ensembles, we may have a better and more flexible control over the spectral property due to the extra degree of freedom in phases.
%%%%%%%%%%%%%%%%%%%%%%%%%%%%%%%%%%%%%%%%%%%%%%%%%%%%%%%%%%%%%%
%%%%%%%%%%%%%%%%%%%%%%%%%%%%%%%%%%%%%%%%%%%%%%%%%%%%%%%%%%%%%%
%%%%%%%%%%%%%%%%%%%%%%%%%%%%%%%%%%%%%%%%%%%%%%%%%%%%%%%%%%%%%%
%%%%%%%%%%%%%%%%%%%%%%%%%%%%%%%%%%%%%%%%%%%%%%%%%%%%%%%%%%%%%%
\begin{figure}[t]
\begin{center}
\includegraphics[height=4.5cm, width=10.0cm]{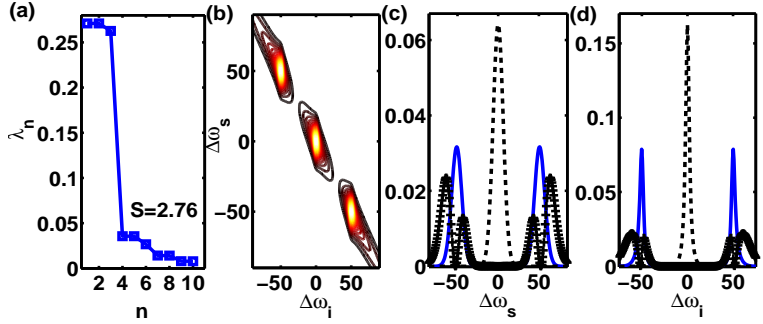}
\caption{Schmidt decomposition and mode probability densities for three multiplexed atomic ensembles with $\theta_{1,2}$ $=$ $0$.\ In (a) and (b) we demonstrate Schmidt eigenvalues and absolute spectral distribution at $\delta p_1/\Gamma_3$ $=$ $50$.\ The associated first two degenerate, $n$ $=$ $1$ and $2$ (solid), and the next two, $n$ $=$ $3$ and $4$ (dash and $+$), mode probability densities of (c) signal $|\psi_n|^2$ and (d) idler photons $|\phi_n|^2$ are plotted accordingly.}\label{fig7}
\end{center}\end{figure}
%%%%%%%%%%%%%%%%%%%%%%%%%%%%%%%%%%%%%%%%%%%%%%%%%%%%%%%%%%%%%%

Finally we investigate the spectral property with zero phases and large frequency shifts in figure \ref{fig7} which has large entropy of entanglement S.\ Pairwise eigenvalues appear again along with a comparable one for every third eigenvalues.\ We also see the degeneracies in mode probability densities which have peaks on the sides to the axis $\Delta\omega_{s,i}$ $=$ $0$.\ For the symmetrical spectral functions we consider here, the S increases as we multiplex more atomic ensembles.\ In the limit of large number of atomic ensembles being multiplexed, we can estimate the S as a combination of the entropy in terms of qudit state of dimensions N$_{\rm MP}$ with the excess entropy due to the entanglement in continuous frequency space, that is S $=$ S$_{d}$ $+$ S$_{\rm N_{\rm MP}=1}$ \cite{Jen2015-2}.\ S$_{\rm N_{\rm MP}=1}$ is the entropy of entanglement for our single biphoton state in frequency space while S$_{d}$ $=$ $\log_2(\rm N_{\rm MP})$ from the conventional qudit state,
\begin{eqnarray}
|\Psi\rangle_d=\sum_{m=1}^{\rm N_{\rm MP}}\frac{1}{\sqrt{\rm N_{\rm MP}}}\hat{a}_{s,m}^\dag\hat{a}_{i,m}^\dag|0\rangle,\label{eq14}
\end{eqnarray}
where $m$ denotes the associated biphoton modes.\ In principle we may generate large entropy of entanglement in high dimensions of photon frequency space from our scheme of the multiplexed atomic ensembles.
\section{Discussions and conclusions}
We propose a scheme that controls the frequency and phase shifts of the cascade emissions from the multiplexed atomic ensembles in which we can manipulate the spectral property of the biphoton state.\ We study the entropy of entanglement S in details for two and three atomic ensembles with dependences on frequency and phase shifts that can be controlled by acoustic-optic modulators and cross-phase modulation experiments respectively.\ We can generate large S by increasing the frequency shift until it saturates, and locate the optimal phases to create the maximal S with finite frequency shifts.\ The extra degrees of freedom in phases other than just frequency shifts provide a fruitful and versatile quantum information control.\ In addition the mode probability densities show double or triple peaks indicating an interference pattern in time domains, which can be measurable and distinguishable from the other modes.\ We would expect of a complexity arising in more than four atomic ensembles in the perspective of optimizing parameters of frequency and phase shifts since it would be harder to map out completely the S in multi-dimensional frequency and phase spaces.\ In principle our scheme opens up a new avenue to entropy control and manipulation in continuous frequency and phase spaces, which can generate large S either by multiplexing more atomic ensembles or by increasing S$_{\rm N_{\rm MP}=1}$, the entropy of entanglement for a single atomic ensemble, which can be done with a larger superradiant decay constant $\Gamma_3^{\rm N}$ or a shorter pulse width $\tau$ \cite{Jen2012-2}.

To multiplex more atomic ensembles in large scale, we may utilize the optical lattices to generate two or three-dimensional arrays of ensembles that can be individually addressed by light-matter interactions.\ In this way even larger S can be created in our scheme to realize high dimensional entanglement \cite{Dada2011, Agnew2011, Fickler2012}.\ This provides a possibility of unlimited communication capacity that is useful in quantum key distribution \cite{Gisin2002} and quantum information application in continuous variables \cite{Braunstein2005}.

For the perspective of experimental measurements of Schmidt eigenvalues or the entropy of entanglement, it requires a technique that operates the mode selection.\ Spectral filtering technique \cite{Babbitt1998} uses transmission gratings with predetermined spectral transfer functions to deflect the input optical pulses such that the spectral information is mapped to the spatial one.\ Similar spectrometer utilizing spectral-to-spatial mapping has been proposed in the planar holographic devices \cite {Mossberg2001} and experimentally demonstrated on a disordered photonic chip \cite{Redding2013}.\ After the spectral calibration, the reconstructed spectra can genuinely retrieve narrow spectral lines or multiple spectral lines with varying amplitudes \cite{Redding2013}.\ The alternative technique of quantum pulse gate \cite{Eckstein2011} uses the sum frequency generation in the setting of parametric down conversion to select out the spectral modes.\ The input state and the shaped pump fields are coupled to the nonlinear waveguide that the selected mode (output) is converted to the sum frequency of both and is separated from the other orthogonal modes.\ It acts effectively as the tomographic reconstruction of the mode characteristics \cite{Brecht2014}.\ The Schmidt eigenvalues can then be retrieved from the probability for the mode selection procedure \cite{Reddy2014} if the quantum efficiency of the conversion is made high enough \cite{Brecht2014}.

Our proposed scheme not only takes advantage of the telecom bandwidth that is favorable in low-loss long-distance quantum communication but also offers an alternative frequency encoding/decoding platform in the multiplexed biphoton state.\ Especially for the modes in well-separated frequency domains, we can individually encode on the spectral property via frequency bins \cite{Bernhard2013} such that Hadamard codes \cite{Lukens2014} for example can be implemented and decoded via coincidence measurements on our signal and idler photons.\ Using the cascade emissions from the multiplexed atomic ensembles, we expect of a potentially efficient entropy manipulation in the biphoton state with controllability and flexibility in conventional quantum optical experiments.\ Our scheme provides an alternatively promising setup in accessing communication capacity in continuous variables, and paves the way toward multimode quantum information processing.
\ack
We acknowledge funding by the Ministry of Science and Technology, Taiwan, under Grant No. MOST-101-2112-M-001-021-MY3 and the support of NCTS on this work.\ We are also grateful for fruitful discussions with S.-Y. Lan and Y.-C. Chen.
%\appendix


\section*{References}

\begin{thebibliography}{100}% quantum repeater
\bibitem{Briegel1998} Briegel H-J, D\"{u}r W, Cirac J I and Zoller P 1998 {\em Phys. Rev. Lett.} {\bf 81} 5932
\bibitem{Dur1999} D\"{u}r W, Briegel H-J, Cirac J I and Zoller P 1999 {\em Phys. Rev. A } {\bf 59} 169
%%%%%%%%%%%%%%%%%%%%%%%%%%%%%%%%%%
\bibitem{Duan2001} Duan L-M, Lukin M D, Cirac J I and Zoller P 2001{\em Nature} {\bf 414} 413
\bibitem{Pirandola2015} Pirandola S, Eisert J, Weedbrook C, Furusawa A and Braunstein S L 2015 {\em Nature Photon.} {\bf 9} 641
\bibitem{Matsukevich2004} Matsukevich D N and Kuzmich A 2004 {\em Science} {\bf 306} 663 
\bibitem{Chou2004} Chou C W, Polyakov S V, Kuzmich A, and Kimble H J 2004 {\em Phys. Rev. Lett.} {\bf 92} 213601
\bibitem{Chaneliere2005} Chaneli\`{e}re T, Matsukevich D N, Jenkins S D, Lan S-Y, Kennedy T A B, and Kuzmich A 2005 {\em Nature } {\bf 438} 833
\bibitem{Chen2006} Chen S, Chen Y-A, Strassel T, Yuan Z-S, Zhao B, Schmiedmayer J, and Pan J-W 2006 {\em Phys. Rev. Lett.} {\bf 97} 173004
\bibitem{Laurat2006} Laurat J, de Riedmatten H, Felinto D, Chou C-W, Schomburg E W, and Kimble H J 2006 {\em Opt. Exp.} {\bf 14} 6912
\bibitem{Chaneliere2006} Chaneli\`{e}re T, Matsukevich D N, Jenkins S D, Lan S-Y, Zhao R, Kennedy T A B and Kuzmich A 2006 {\em Phys. Rev. Lett.} {\bf 97} 093604
\bibitem{Radnaev2010} Radnaev A G, Dudin Y O, Zhao R, Jen H H, Jenkins S D, Kuzmich A, and Kennedy T A B 2010 {\em Nature Physics} {\bf 6} 894
\bibitem{Jen2010} Jen H H and Kennedy T A B 2010 {\em Phys. Rev. A} {\bf 82} 023815
\bibitem{Jen2012-2} Jen H H 2012 {\em J. Phys. B: At. Mol. Opt. Phys.} {\bf 45} 165504
%%%%%%%%%%%%%%%%%%%%%% discrete polarization and frequency qubits Experiment
\bibitem{Clauser1969} Clauser J F, Horne M A, Shimony A, and Holt B A 1969 {\em Phys. Rev. Lett.} {\bf 23}, 880
\bibitem{Aspect1981} Aspect A, Grangier P, and Roger G 1981 {\em Phys. Rev. Lett.} {\bf 47}, 460
\bibitem{Kwiat1995} Kwiat P G, Mattle K, Weinfurter H, Zeilinger A, Sergienko A V, and Shih Y 1995 {\em Phys. Rev Lett.} {\bf 75}, 4337
\bibitem{Lan2007} Lan S-Y, Jenkins S D, Chaneli\`{e}re T, Matsukevich D N, Campbell C J, Zhao R, Kennedy T A B, and Kuzmich A 2007 {\em Phys. Rev. Lett.} {\bf 98} 123602
\bibitem{Ramelow2009} Ramelow S, Ratschbacher L, Fedrizzi A, Langford N K, and Zeilinger A 2009 {\em Phys. Rev. Lett.} {\bf 103}, 253601
\bibitem{Gisin2002} Gisin N, Ribordy G, Tittel W, and Zbinden H 2002 {\em Rev. Mod. Phys.} {\bf 74}, 145
\bibitem{Braunstein2005} Braunstein S L and Van Loock P 2005 {\em Rev. Mod. Phys.} {\bf 77}, 513
%%%%%%%% Schmidt decomposition, transverse momentum, space and frequency entanglement, CV
\bibitem{Branning1999} Branning D, Grice W P, Erdmann R, and Walmsley I A 1999 {\em Phys. Rev. Lett.} {\bf 83}, 955
\bibitem{Law2000} Law C K, Walmsley I A, and Eberly J H 2000 {\em Phys. Rev. Lett.} {\bf 84}, 5304
\bibitem{Parker2000} Parker S, Bose S, and Plenio M B 2000 {\em Phys. Rev. A} {\bf 61}, 032305
\bibitem{Law2004} Law C K and Eberly J H 2004 {\em Phys. Rev. Lett.} {\bf 92}, 127903
\bibitem{Moreau2014} Moreau P-A, Devaux F, and Lantz E 2014 {\em Phys. Rev. Lett.} {\bf 113}, 160401
\bibitem{Grad2012} Grodecka-Grad A, Zeuthen E, and S\o rensen A S 2012 {\em Phys. Rev. Lett.} {\bf 109}, 133601
%%%%%%%%%%%%%%%%%%%%%%%%%%%%%%%%% OAM
\bibitem{Arnaut2000} Arnaut H H and Barbosa G A 2000 {\em Phys. Rev. Lett.} {\bf 85}, 286
\bibitem{Mair2001} Mair A, Vaziri A, Weihs G, and Zeilinger A 2001 {\em Nature} {\bf 412}, 313
\bibitem{Molina2007} Molina-Terriza G, Torres J P, and Torner L 2007 {\em Nature Phys.} {\bf 3}, 305
%%%%%%%%%%%%%%%%%%%%%% high OAM
\bibitem{Dada2011} Dada A C, Leach J, Buller G S, Padgett M J, and Andersson E 2011 {\em Nat. Phys.} {\bf 7}, 677
\bibitem{Agnew2011} Agnew M, Leach J, McLaren M, Roux F S, and Boyd R W 2011 {\em Phys. Rev. A} {\bf 84}, 062101
\bibitem{Fickler2012} Fickler R, Lapkiewicz R, Plick W N, Krenn M, Schaeff C, Ramelow S, Zeilinger A 2012 {\em Science} {\bf 338}, 640
%%%%%%%%%%%%%%%%%%%%%% quantum storage of OAM
\bibitem{Nicolas2014} Nicolas A, Veissier L, Giner L, Giacobino E, Maxein D, and Laurat J 2014 {\em Nat. Photonics} {\bf 8}, 234
\bibitem{Ding2015} Ding D-S, Zhang W, Zhou Z-Y, Shi S, Xiang G-Y, Wang X-S, Jiang Y-K, Shi B-S, and Guo G-C 2015 {\em Phys. Rev. Lett.} {\bf 114}, 050502
%%%%%%%%%%%%%%%%%%%%%% Frequency combs memory, Multiplex
\bibitem{Afzelius2009} Afzelius M, Simon C, De Riedmatten H, and Gisin N 2009 {\em Phys. Rev. A} {\bf 79}, 052329
\bibitem{Zheng2015} Z. Zheng, O. Mishina, N. Treps, and C. Fabre, Atomic quantum memory for multimode frequency combs, Phys. Rev. A {\bf 91}, 031802(R) (2015).
\bibitem{Collins2007} Collins O A, Jenkins S D, Kuzmich A, and Kennedy T A B 2007 {\em Phys. Rev. Lett.} {\bf 98}, 060502
\bibitem{Lan2009} Lan S-Y, Radnaev A G, Collins O A, Matsukevich D N, Kennedy T A B and Kuzmich A 2009 {\em Opt. Exp.} {\bf 17}, 13639
\bibitem{Simon2007} Simon C, De Riedmatten H, Afzelius M, Sangouard N, Zbinden H, and Gisin N 2007 {\em Phys. Rev. Lett.} {\bf 98}, 190503
%%%%%%%%%%%%%%%%%%%%%% spectral shaping Experiment
\bibitem{Bernhard2013} Bernhard C, Bessire B, Feurer T, and Stefanov A 2013 {\em Phys. Rev. A} {\bf 88}, 032322
\bibitem{Lukens2014} Lukens J M, Dezfooliyan A, Langrock C, Fejer M M, Leaird D E, and Weiner A M 2014 {\em Phys. Rev. Lett.} {\bf 112}, 133602
\bibitem{Jen2015-2} Jen H H 2015 {\em arXiv:1510.01859}
%%%%%%%%%%%%%%%%%%%%%% cascade emissions of SR and CLS
\bibitem{QO:Scully} Scully M O and Zubairy M S 1997 {\em Quantum Optics} (Cambridge University Press)
\bibitem{Eberly2006} Eberly J H 2006 {\em J. Phys. B: At. Mol. Opt. Phys.} {\bf 39}, S599
\bibitem{Scully2006} Scully M O, Fry E S, Raymond Ooi C H, and W\'{o}dkiewicz K 2006 {\em Phys. Rev. Lett.} {\bf 96}, 010501
\bibitem{Mazets2007} Mazets I E and Kurizki G 2007 {\em J. Phys. B: At. Mol. Opt. Phys.} {\bf 40} F105
\bibitem{Jen2012} Jen H. H. 2012 {\em Phys. Rev. A} {\bf 85}, 013835
\bibitem{Srivathsan2014} Srivathsan B, Gulati G K, Chng B, Maslennikov G, Matsukevich D, and Kurtsiefer C 2014 {\em Phys. Rev. Lett.} {\bf 111}, 123602
\bibitem{Lehmberg1970} R H Lehmberg 1970 {\em Phys. Rev. A} {\bf 2}, 883
\bibitem{Dicke1954} R H Dicke 1954 {\em Phys. Rev} {\bf 93}, 99
\bibitem{Rehler1971} Rehler N E and Eberly J H 1971 {\em Phys. Rev. A} {\bf 3} 1735
\bibitem{Friedberg1973} Friedberg R, Hartmann S R, and Manassah J T 1973 {\em Phys. Rep.} {\bf 7}, 101
\bibitem{Scully2009} Scully M O 2009 {\em Phys. Rev. Lett.} {\bf 102}, 143601
\bibitem{Jen2015} Jen H H 2015 {\em Annals of Phys.} {\bf 360}, 556
\bibitem{Kang2003} Kang H and Zhu Y 2003 {\em Phys. Rev. Lett.} {\bf 91}, 093601
\bibitem{Lo2010} Lo H-Y, Su P-C and Chen Y-F 2010 {\em Phys. Rev. A} {\bf 81}, 053829
\bibitem{Shiau2011} Shiau B-W, Wu M-C, Lin C-C and Chen Y-C 2011 {\em Phys. Rev. Lett.} {\bf 106}, 193006
\bibitem{Venkataraman2013} Venkataraman V, Saha K, and Gaeta A L 2013 {\em Nature Photon.} {\bf 7}, 138
\bibitem{Babbitt1998} Babbitt W R and Mossberg T W 1998 {\em Optics Comm.} {\bf 148}, 23 
\bibitem{Mossberg2001} Mossberg T W 2001 {\em Optics Lett.} {\bf 26}, 414
\bibitem{Redding2013} Redding B, Liew S F, Sarma R, and Cao H 2013 {\em Nature Photon.} {\bf 7}, 746
\bibitem{Eckstein2011} Eckstein A, Brecht B, and Silberhorn C 2011 {\em Optics Exp.} {\bf 19}, 13770
\bibitem{Brecht2014} Brecht B, Eckstein A, Ricken R, Quiring V, Suche H, Sansoni L, and Silberhorn C 2014 {\em Phys. Rev. A} {\bf 90}, 030302(R)
\bibitem{Reddy2014} Reddy D V, Raymer M G, and McKinstrie C J 2014 {\em Optics Lett.} {\bf 39}, 2924
\end{thebibliography}
\end{document}